\documentclass[10pt, conference, compsocconf]{IEEEtran}
\usepackage{amsmath}
\usepackage[T1]{fontenc}

\usepackage{amsfonts}
\usepackage{amssymb}
\usepackage[normalem]{ulem}
\usepackage[hyphens]{url}
\usepackage[sort,nocompress]{cite}
\usepackage[final]{microtype}
\usepackage[keeplastbox]{flushend}
\newcommand{\ignore}[1]{}
\usepackage[pass]{geometry}
\usepackage{fancyhdr}
\usepackage{color}
\usepackage{soul}
\usepackage{bm}
\usepackage{blindtext}
\usepackage{booktabs}
\usepackage{listings}
\usepackage{xcolor}
\usepackage{bold-extra}
\usepackage[scaled=0.825]{beramono}
\usepackage[export]{adjustbox}
\newcommand{\name}{$A^3$ }
\newcommand{\namet}{{\large ${A^3}$}~}
\newcommand{\namens}{$A^3$}
\newcommand{\para}[1]{\noindent \textbf{#1.}}

\graphicspath{{figures/}}
\usepackage{multirow, makecell}
\newcolumntype{P}[1]{>{\centering\arraybackslash}p{#1}}
\newcolumntype{L}[1]{>{\raggedright\let\newline\\\arraybackslash\hspace{0pt}}m{#1}}
\newcolumntype{C}[1]{>{\centering\let\newline\\\arraybackslash\hspace{0pt}}m{#1}}
\newcolumntype{R}[1]{>{\raggedleft\let\newline\\\arraybackslash\hspace{0pt}}m{#1}}
\newcommand{\codesize}{\fontsize{8.0pt}{8.0pt}\selectfont}
\lstdefinestyle{myStyle}{
  belowcaptionskip=1\baselineskip,
  frame=tb,
  language=c++,
  aboveskip=0mm,
  belowskip=0mm,
  showstringspaces=false,
  columns=flexible,
  basicstyle={\codesize\fontfamily{fvm}\selectfont},
  numbers=left,
  xleftmargin=3em,
  numberstyle={\color{gray}\texttt},
  keywordstyle=\color{black}\textbf,
  commentstyle=\color{gray},
  stringstyle=\color{mauve},
  frame=none,
  breaklines=true,
  breakatwhitespace=true,
  tabsize=2,
  morekeywords={MatrixMult, Softmax, parfor, parallel, each, not, in},
  deletekeywords={and},
}

\let\labelindent\relax

\usepackage{enumitem}
\setlist{noitemsep, leftmargin=*, topsep=0pt, partopsep=0pt}
\renewcommand{\IEEEbibitemsep}{0.2pt plus 0.5pt}
\makeatletter
\IEEEtriggercmd{\reset@font\normalfont\fontsize{7.0pt}{7.2pt}\selectfont}
\makeatother
\IEEEtriggeratref{1}

\begin{document}
\title{A$\Large\bm{^3}$: Accelerating Attention Mechanisms in Neural Networks with Approximation\vspace{-0.1in}}
\author{\fontfamily{phv}\fontsize{10}{10} \selectfont \hspace{-0.1in}Tae Jun Ham*~~~~~~~~Sung Jun Jung*~~~~~~~~Seonghak Kim*~~~~~~~~~Young H. Oh$^{\dagger}$~~~~~~~Yeonhong Park*\\\vspace{-0.07in}\\
\fontfamily{phv}\fontsize{10}{10} \selectfont \hspace{-0.1in}Yoonho Song*~~~~Jung-Hun Park*~~~~Sanghee Lee*~~~~Kyoung Park$^{\ddagger}$~~~~Jae W. Lee*~~~~Deog-Kyoon Jeong*\\\vspace{0.025in}\\
\fontfamily{phv}\fontsize{10}{10} \selectfont *Seoul National University~~~~~~~$^{\dagger}$Sungkyunkwan University~~~~~~~$^{\ddagger}$SK Hynix\\\vspace{-0.05in}\\
\fontsize{9}{9}\selectfont \texttt{\{taejunham, miguel92, ksh1102, ilil96, yhsong, jhpark94, shlee95, jaewlee, dkjeong\}@snu.ac.kr},\\\vspace{-0.15in}\\ 
\fontsize{9}{9}\selectfont \texttt{younghwan@skku.edu, kyoung.park@sk.com}\vspace{-0.1in}
}
% Intentionally updated Yoonho Song & Jung-Hun Park's names

% %\author{\IEEEauthorblockN{Authors Name/s per 1st Affiliation (Author)}
% \IEEEauthorblockA{line 1 (of Affiliation): dept. name of organization\\
% line 2: name of organization, acronyms acceptable\\
% line 3: City, Country\\
% line 4: Email: name@xyz.com}
% \and
% \IEEEauthorblockN{Authors Name/s per 2nd Affiliation (Author)}
% \IEEEauthorblockA{line 1 (of Affiliation): dept. name of organization\\
% line 2: name of organization, acronyms acceptable\\
% line 3: City, Country\\
% line 4: Email: name@xyz.com}
% }
\maketitle

\begin{abstract}
With the increasing computational demands of neural networks, many hardware accelerators for the neural networks have been proposed. Such existing neural network accelerators often focus on popular neural network types such as convolutional neural networks (CNNs) and recurrent neural networks (RNNs); however, not much attention has been paid to \emph{attention mechanisms}, an emerging neural network primitive that enables neural networks to retrieve most relevant information from a knowledge-base, external memory, or past states. The attention mechanism is widely adopted by many state-of-the-art neural networks for computer vision, natural language processing, and machine translation, and accounts for a large portion of total execution time. We observe today's practice of implementing this mechanism using matrix-vector multiplication is suboptimal as the attention mechanism is semantically a content-based search where a large portion of computations ends up not being used. Based on this observation, we design and architect \namens, which accelerates attention mechanisms in neural networks with algorithmic \emph{approximation} and hardware \emph{specialization}. Our proposed accelerator achieves multiple orders of magnitude improvement in energy efficiency (performance/watt) as well as substantial speedup over the state-of-the-art conventional hardware. 
\end{abstract}

\IEEEpeerreviewmaketitle

% Contents
\section{Introduction} \label{sec:intro}
Neural networks (NNs) are currently one of the most popular techniques to perform complex AI tasks in various domains such as computer vision, natural language processing, robotics, etc. With a large amount of data, NNs can effectively solve a wide range of AI challenges and can often surpass human performance in many domains. However, these advantages of NNs come at a high computational cost involving tens of billions of floating-point operations. %For example, VGG-16 \cite{vgg16}, one of the most complex NN models used for image classification, requires 15.5 billion multiply and add operations to process a 224$\times$224 image \cite{szetutorial}. 
In order to minimize the energy cost of such a large number of operations and maximize the throughput of NN processing, many prior works proposed various FPGA or ASIC based accelerators \cite{diannao, eyeriss, tpu}. These prior works are indeed very effective in improving the performance and energy efficiency of the popular NN types such as convolutional neural networks (CNNs) or recurrent neural networks (RNNs). However, such accelerators do not provide full support for an emerging NN primitive such as attention mechanisms. 

Attention mechanisms are one of the most important recent advancements in neural networks. Unlike CNNs and RNNs, which has a limited capacity to utilize information from the past or external knowledge, attention mechanisms (also called memory mechanisms) enable NNs to access and utilize such information by providing extra connections to past state cells, explicit memory cells, and so on. Naturally, not all information from past state cells or explicit memory cells is equally relevant to what NN is currently processing. Hence, the attention mechanism determines what is relevant to the currently processed information through content-based similarity search and decides where to {\it attend}. This attention mechanism has been recently adopted in many domains of NNs such as computer vision \cite{showattend, movieqa, rwmn} and natural language processing \cite{transformer, memn2n, bert, nmt}. In addition, this mechanism is also used to enable NNs to solve a class of complex problems that were previously difficult for conventional NNs due to their lack of ability to memorize and retrieve data \cite{ntm, dnc}. 

In the conventional hardware, an attention mechanism is usually implemented as dense matrix operations and softmax operations. A dense matrix-vector multiplication operation computes the similarity across all search targets and thus the computational complexity of this operation is proportional to the number of search targets. In other words, attention mechanism requires more computation when a NN model wants to retrieve relevant information over the larger external knowledge-base, over a longer period of past information, or from a longer sequence of data. To make it even worse, in some NN models utilizing self-attention mechanisms~\cite{transformer, decomatt, strself}, the computational complexity of attention mechanism is proportional to the square of the search targets (e.g., a length of the reading passage in the reading comprehension task). Naturally, this large computational cost of attention mechanism becomes a limiting factor for the capacity of the NN models and accounts for a significant portion of the performance and energy cost of existing models.

To address this challenge and mitigate the bottlenecks arising from the computational cost of attention mechanisms, we architect \namens, a hardware accelerator for attention mechanisms in NNs. To design an efficient accelerator, we not only focus on the efficient implementation of the attention mechanism in hardware but also focus on reducing the amount of computation in attention mechanism through algorithmic optimization and approximation. In particular, based on the observation that not all search targets are equally likely to be relevant, our design presents an approximate candidate selection mechanism to reduce the number of search targets, and thus the amount of computation. Furthermore, we propose a specialized hardware pipeline exploiting parallelism to accelerate approximated attention mechanisms while making it even more efficient. With this algorithm-hardware co-design, our proposed accelerator offers significant performance improvements and orders of magnitudes improvements in energy efficiency (performance/W), thereby enabling existing NN models with attention mechanisms to utilize larger external knowledge or a longer sequence of data. Our contributions are summarized as follows.

\begin{itemize}
    \item We quantify the attention mechanism bottlenecks in NN models and identify that a substantial portion of time in NN models is spent on attention mechanisms. 
    \item We exploit the potential for approximation in attention mechanisms and present an approximation scheme which enables our proposed hardware to find potentially relevant search targets while avoiding an exhaustive search. 
    \item We design a specialized hardware pipeline for an attention mechanism exploiting parallelism and datapath specialization to significantly improve the performance and energy efficiency of the attention mechanism. 
    \item We demonstrate that our proposed accelerator achieves multiple orders of magnitude speedup and energy efficiency over conventional hardware. Furthermore, with our specialized hardware for the approximation scheme, \name achieves even higher speedup and energy efficiency while minimizing the degradation of the model accuracy. 
\end{itemize}
\section{Background and Motivation} \label{sec:motiv}
\subsection{Attention Mechanism}

\begin{figure}[t]
\begin{minipage}[b]{1.02\hsize}
\begin{lstlisting}[escapechar=\^,style=myStyle,mathescape=true]
// key: $n \times d$, value: $n \times d$, query, output: $d$
float[] $\textbf{attention\_mechanism}$ (float key[][],
  float value[][], float query[]) {
  /* Step 1 : Dot-Product Computation */
  for i = 0 to n:
    sum = 0
    for j = 0 to d:
      sum += key[i][j] * query[j]
    dot_product[i] = sum
  /* Step 2 : Softmax Computation */
  score = softmax(dot_product)
  /* Step 3 : Output Computation */
  for j = 0 to d:
    sum = 0
    for i = 0 to n:
      sum += score[i] * value[i][j]
    output[j] = sum
  return output
}
float[] $\textbf{softmax}$(float input[]) {
  sum = 0
  for i = 0 to n:
    sum += exp(input[i])
  for i = 0 to n:
    output[i] = exp(input[i]) / sum
  return output
}
\end{lstlisting}
\end{minipage}
\caption{Pseudocode for an attention mechanism.}
\label{fig:algorithm}
\end{figure}

\para{Operation} Attention mechanism is essentially a content-based search. Figure \ref{fig:algorithm} represents the computation of the attention mechanism in pseudocode. Given a query vector with $d$ dimensions and a key matrix with $n$ vectors where each vector has $d$ dimensions, the attention mechanism first computes a similarity score (i.e., dot-product) for each entry in the key matrix (Step 1 in Figure \ref{fig:algorithm}). After this process, an $n$-dimensional vector (i.e., {\tt dot\_product[]}) is obtained. This array is then processed with softmax function (Step 2 in Figure \ref{fig:algorithm}). Finally, the normalized score is used as a weight to retrieve the weighted sum of vectors from the $n \times d$ value matrix (Step 3 in Figure \ref{fig:algorithm}). In short, the set of indices for the set of vectors in the key matrix which are similar to the query vector is first obtained, and these indices (along with weight values) are used to obtain the weighted sum for the set of vectors from the value matrix. This mechanism is often called {\it soft} attention mechanism since it only consists of differentiable computations, which makes this mechanism well-suited for NNs which are trained with back-propagation. 
% \footnote{Technically, this $d$ can be different from that of the key matrix but in practice, they are often the same and thus we focus on the case where both matrices have the same number of columns.} 
\begin{figure}[t]
  \centering
  \includegraphics[width=.9\linewidth]{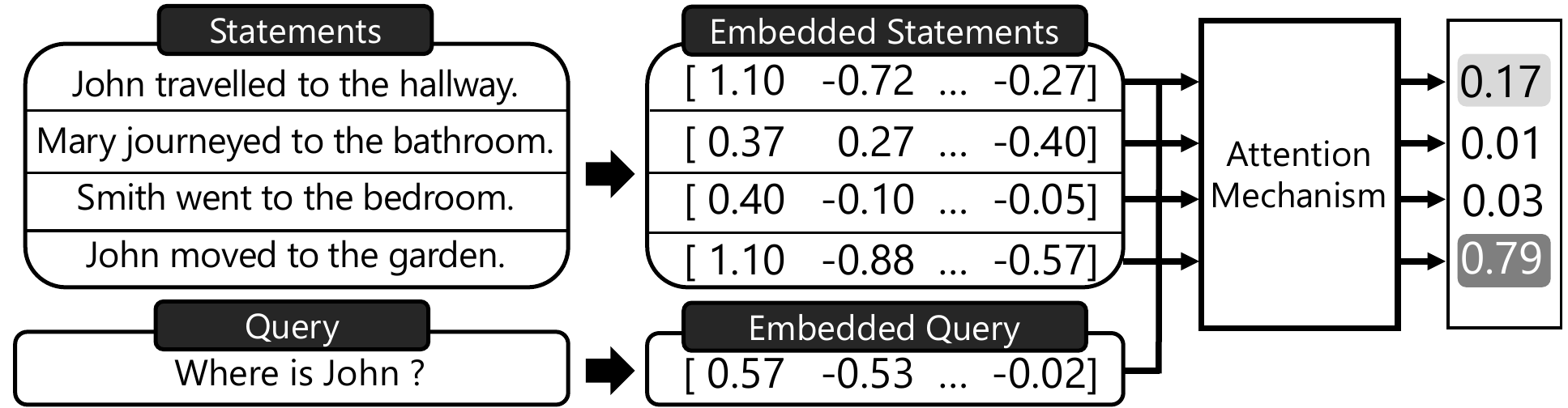} 
  \caption{Example application of attention mechanism (Step 1 and 2 in Figure~\ref{fig:algorithm}) from Facebook bAbI QA \cite{babiqa}.}
  \label{fig:example}
\end{figure}

\para{Example Application} Figure \ref{fig:example} introduces a very simple example which shows how the attention mechanism is utilized to enable a NN to find a sentence that is relevant to the question in the Facebook bAbI QA task \cite{babiqa}. In this task, a list of statements and a question are provided in a natural language. The goal is to find the right answer to the question. Note that not all provided statements are necessary to answer a question. In many NN models solving this task, an attention mechanism is utilized to identify the most relevant statements among these provided statements. For example, End-to-End Memory Network \cite{memn2n} first embeds (i.e., converting natural language into vector representation as in Word2Vec~\cite{word2vec}, Glove~\cite{glove}, FastText~\cite{fasttext}) each statement sentence and query sentence. Then, using the attention mechanism, it finds the most relevant sentence to the query from the set of statements. For example, as shown in Figure \ref{fig:example}, the attention mechanism can identify that ``{\it John moved to the garden.}'' is the most relevant sentence for a query ``{\it Where is John?}'' by performing a similarity search using the embeddings. If multiple sentences are required to answer the question, it updates the query with the relevant sentence found in the previous iteration and utilizes the attention mechanism again to retrieve other relevant sentences from the set. After obtaining all relevant sentences, it utilizes a final weight matrix to generate the final answer.  

\subsection{Cost of Attention Mechanism} 

For a given $n$ and $d$, an attention operation requires i) $nd$ multiplications and $n(d-1)$ additions in Step 1, ii) $n$ exponent computations and $n-1$ additions, and $n$ divisions in Step 2, and iii) $nd$ multiplications and $(n-1)d$ additions in Step 3. Naturally, the number of these computations increases almost-linearly with both $n$ and $d$. Here, $n$ represents the number of data in an external knowledge base or the number of past states that this mechanism allows models to look for. Naturally, the larger $n$ allows more powerful NN models. On the other hand, $d$ is the dimension of a single vector entry. A single vector in the key matrix usually represents an embedding of a word, a sentence, a knowledge, or any other portion of a larger entity. The larger $d$ allows this embedding to have a richer space and thus can provide a higher quality embedding. Between $n$ and $d$, $n$ is much more likely to increase further since a larger $n$ directly makes the NN models more powerful by allowing it to search over a larger number of data to extract useful information. 

We analyze the portion of the time spent on attention mechanisms in popular NN models. For this analysis, we focus on three different workloads utilizing attention mechanisms: End-to-End Memory Network (MemN2N)\cite{memn2n} running bAbI QA task \cite{babiqa}, Key-Value Memory Network (KV-MemN2N) \cite{memkvnn} running WikiMovies QA task \cite{memkvnn}, and Google BERT \cite{bert} running SQuAD task \cite{squad} (see Section~\ref{sec:workload} for details). We run these workloads on Intel Xeon Gold 6128 CPU (all workloads) and NVIDIA Titan V GPU (BERT only) then report the profiled data below. 

\begin{figure}[!t]
  \centering
  \includegraphics[width=0.9\linewidth]{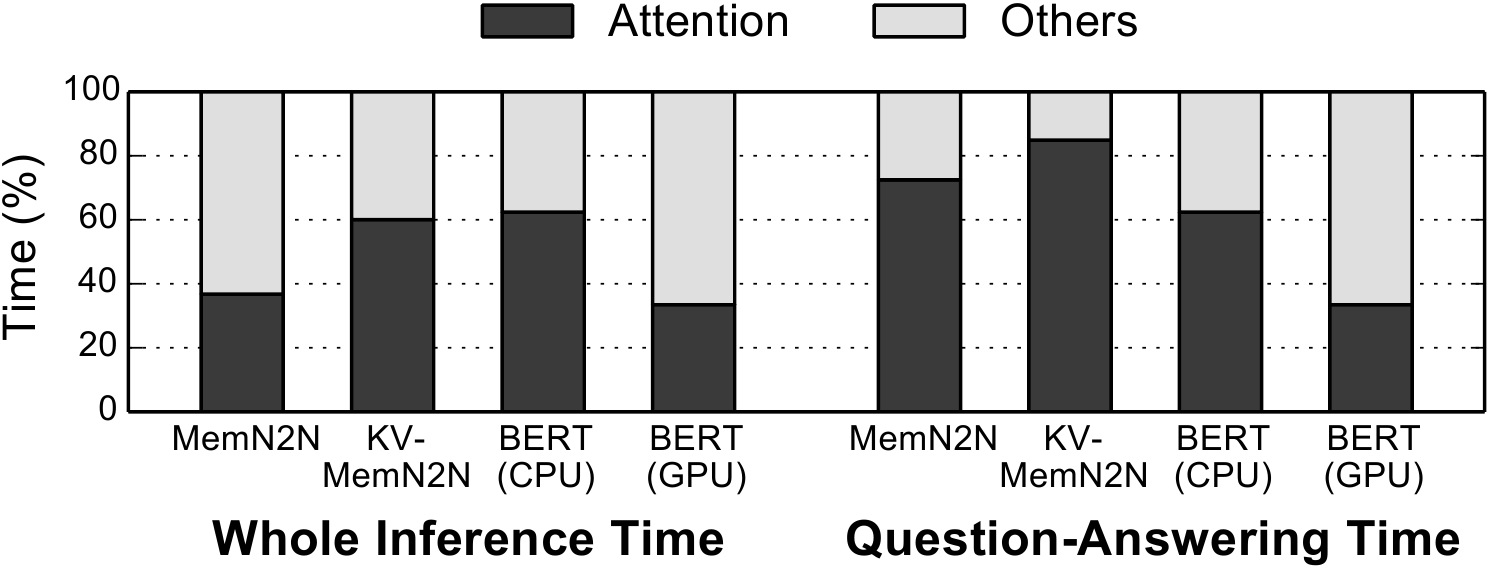}
  \caption{Portion of the time accountable for attention mechanism of NN workloads (for the total inference time and for the query response time).}
  \label{fig:motiv}
\end{figure}

\begin{figure*}[t]
  \centering
  \includegraphics[width=0.85\textwidth, trim={0 0pt 0 0},clip] {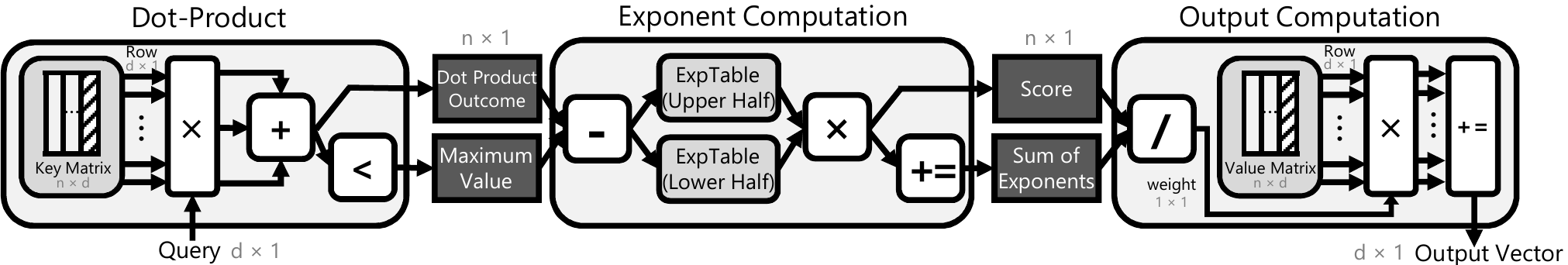}
  \vspace{-0.15in}
  \caption{Block diagram of the base \name}
  \label{fig:hwblock}
\end{figure*}
Figure~\ref{fig:motiv} shows that the attention mechanism is responsible for the significant portion (i.e., over 35\% in all workloads) of the inference runtime. If we take a more in-depth look into the nature of these tasks, the actual cost of the attention model is even higher. Most models handling QA tasks take a substantial amount of time on comprehension (e.g., embedding generation) of the provided knowledge (e.g., list of statements in bAbI QA task or Wikipedia-oriented information about various movies in WikiMovies task). Those processes are query-independent and thus it is possible to preprocess them before a query is provided. Thus, the actual critical path (i.e., query response time) of the question-answering task often does not include such time. The right side of Figure~\ref{fig:motiv} shows query response time (as opposed to the total inference time which includes both comprehension time and query response time) for all workloads. Compared to the result for the total inference time, the attention mechanism takes noticeably a larger portion of time (over 70\%) in both MemN2N and KV-MemN2N. The portion of the attention mechanism in the BERT model remains the same since it performs comprehension and query response in an integrated manner. To accelerate attention mechanism and make it more energy-efficient, we design a specialized hardware accelerator for the attention mechanism. Section \ref{sec:design} presents how our design efficiently handles this operation.

\subsection{Opportunity for Approximation}

Typically, in many popular NN model implementations, the attention mechanism is implemented as a matrix-vector multiplication (i.e., multiplication of the key matrix and the query vector) followed by a softmax function again followed by another matrix-vector multiplication  (i.e., multiplication of the value matrix and the weight vector). Such implementations can utilize the fast matrix-vector multiplication capability provided by popular NN processing frameworks such as TensorFlow \cite{tensorflow} and Torch \cite{torch}. A dense matrix-vector multiplication-based implementation is functionally correct. However, the nature of this operation is a search, not a dense computation. In reality, most of the computations performed in the first matrix-vector multiplication have very little impact on the final output since most score values become near-zero after the softmax normalization which is essentially a continuous, differentiable approximation of the {\it argmax} operation which selects the index for the maximum number in an array.  

The intuition behind our approximation proposal is to avoid such unnecessary computation. By preprocessing the key matrix, it is possible to obtain a set of candidate rows which are likely to have high score values without much computation. By doing so, our proposed scheme can avoid unnecessary dot product computation, softmax computation, and final result computation. Section \ref{sec:approximation} presents our proposed approximation scheme and Section \ref{sec:approxhw} presents the hardware accelerator module necessary for this approximated attention mechanism.  
\section{\namet Base Design} \label{sec:design}

We introduce the base design of \namens, a specialized hardware accelerator for attention mechanisms in neural networks which can be integrated to either CPU, GPU, or an existing hardware accelerator. \name accelerates the attention mechanism shown in Figure \ref{fig:algorithm} of Section \ref{sec:motiv}. For high throughput and energy efficiency, \name employs a pipeline designed with customized datapath exploiting parallelism. This section provides an in-depth overview of the base design of \name without approximation, and later sections introduce an approximate attention mechanism and extended hardware designs.

\subsection{Pipeline Design}

Base \name takes three inputs --- a key matrix ($n \times d$), a value matrix ($n \times d$), and a query vector ($d$) --- to compute a $d$-dimensional output vector. Figure~\ref{fig:hwalgorithm} shows the computation pattern of the Base \name in pseudocode. Note that this is slightly different from the pseudocode in Figure \ref{fig:algorithm} as we reorder some computations to make it more suited for hardware implementation. Figure~\ref{fig:hwblock} shows the block diagram of the base \name pipeline, which implements the pseudocode in Figure~\ref{fig:hwalgorithm}.  As shown in both figures, the base \name pipeline consists of three modules: i) dot-product module, ii) exponent computation module, and iii) output computation module. Below, we explain each module in greater details. 

\begin{figure}[!th]
\begin{minipage}[b]{1.02\hsize}
\begin{lstlisting}[escapechar=\^,style=myStyle,mathescape=true]
float[] $\textbf{attention\_mechanism}$ (float key[][],
  float value[][], float query[]) {
  /* Module 1: Dot-Product*/
  max = $-\infty$
  for i = 0 to n:
    parallel for j = 0 to d:
      temp[i][j] = key[i][j] * query[j]
    dot_product[i] = ParallelSum(temp[i])
    if dot_product[i] > max:
      max = dot_product[i]
  /* Module 2: Exponent Computation */
  expsum = 0
  for i = 0 to n:
    dot_product[i] -= max
    score[i] = exp(dot_product[i])
    expsum += score[i]
  /* Module 3: Output Computation */
  for i = 0 to n:
    weight[i] = score[i]/expsum
    parallel for j = 0 to d:
      output[j] += weight[i] * value[i][j]
}
\end{lstlisting}
\end{minipage}
\caption{Base \name pipeline represented with pseudocode.}
\label{fig:hwalgorithm}
\end{figure}

\para{Module 1: Dot-Product Computation} The first module of the pipeline (Line 3-10 in Figure~\ref{fig:hwalgorithm}) computes an inner product between a single row of the key matrix and a query vector. This hardware consists of $d$ multipliers and a $d$-way adder tree for a sum reduction operation. For each cycle, a row of the key matrix is loaded (in sequential order) and each of its vector elements is multiplied by the corresponding element of the query vector using the array of multipliers. Then, once the multiplication finishes, these set of values are passed to the adder tree for a parallel sum reduction. The result of this operation is stored in the corresponding register in the dot product outcome register file. This process is repeated for $n$ times until all rows in the key matrix are processed. Note that this module also finds the maximum value among all elements of the {\tt dot\_product} array (Line 9-10 in Figure~\ref{fig:hwalgorithm}). This value is used by the next module. 

\para{Module 2: Exponent Computation} Exponent computation (Line 11-16 in Figure~\ref{fig:hwalgorithm}) computes the exponent of each dot-product value computed by the previous module. Normally, to perform an exponent computation, a hardware exponent computation unit should be used. Instead, our design implements an exponent function using a lookup table. However, there are two challenges with this approach: handling an overflow and reducing the size of the lookup table. First, the outcome of an exponent function increases very rapidly with the increase in input value, and thus it can easily cause an overflow in a fixed-point representation for large input. To counter this, we leverage the fact that softmax is \emph{invariant} to the addition (or subtraction) of a constant to all elements in the vector. Before performing an exponent computation, this module first starts with subtracting the maximum value of the input vector from the dot-product value being processed. With this subtraction, all elements in the input vector will be less than or equal to zero, and thus the outputs of the exponent function will always be less than or equal to 1. 

The second challenge is that the size of the lookup table becomes very large when high precision is required. For example, if we use a 16-bit representation, a lookup table requires 65,536 entries and such an SRAM can incur large power and area overhead. To reduce the size of the lookup table, we exploit the fact that a single exponent operation can be decomposed into a multiplication of two exponent operations. For example, as shown below, computing an exponent for 8-bit input is equivalent to computing an exponent for two 4-bit inputs (i.e., one for the upper 4-bits and another for the lower 4-bits) and multiplying them. 
$$\vspace{-0.04in} e^{0.10101111_2}  = e^{0.10100000_2} \times e^{0.00001111_2} \vspace{-0.04in}$$
With this transformation, instead of building a large lookup table (e.g., one with 65,536 entries), we can utilize two smaller lookup tables (e.g., each with 256 entries) and a multiplier to obtain the same outcome. Once the exponent of the dot-product is computed, this value gets accumulated. This accumulated value is later used as the softmax denominator (Line 19 in Figure~\ref{fig:hwalgorithm}).

\para{Module 3: Output Computation} This module (Line 17-21 in Figure \ref{fig:hwalgorithm}) computes the output of the attention operation. Every cycle, an element of the score vector is divided by the sum of all score values for normalization. Then, this value (i.e., {\tt weight}) serves as a scaling factor for the corresponding row vector in the value matrix. Each element of a row vector is multiplied by this value. The result of this computation is accumulated in the {\tt output} register. This process is repeated for $n$ times. 

\para{Throughput and Latency} Our proposed hardware can handle three queries at a time in a pipelined manner. When a query finishes its computation for a module, it is then passed to the next hardware module, and thus the next query can enter the current module. To balance this pipeline, we deliberately design all three modules to have the matching throughput (i.e., each module takes $n$ cycles + $\alpha$ to process a query). Among three modules, the last module has the longest latency of $n + 9$ ($n$ cycles to handle $n$ rows in a pipelined way, 7 cycles for a division, and 2 cycles for a multiplication and accumulate). Thus, the pipeline latency is $3n+27$ cycles and the throughput is $n+9$ cycles per query.

%As explained in the previous paragraph, our dot-product computation module finds the maximum element among {\tt dot\_product} vector. This module first starts with subtracting the maximum value from a dot product outcome. 

\subsection{Quantization} \label{sec:quantization}

Most NN tasks are tolerant to certain levels of errors by nature and can thus operate with lower bitwidth representations while keeping the accuracy of higher bitwidth representations. In such a case, utilizing a lower bitwidth is crucial to the accelerator design since many of the registers' or computation unit's energy cost scales linearly or quadratically with the bitwidth of a representation. In addition, since floating-point operations cost much more energy than fixed-point ones, it is often beneficial for specialized accelerators to utilize a fixed-point representation. Our model first quantizes the provided floating-point input to $i$ integer bits and $f$ fraction bits (plus a sign bit), and then utilizes different bitwidths for each stage of the pipeline to maintain the precision and avoid an overflow while minimizing the energy cost. Below, we explain our rationale for our choice of fixed-point bitwidths for the pipeline. We evaluate the impact of quantization in Section \ref{sec:eval:accuracy}. 

\para{Integer Bitwidths} Integer bitwidths are determined by the dynamic range of values. Suppose $i$ (e.g., $i=4$ is used for our evaluation) integer bits are required to represent the elements of the input key matrix and the query vector. In other words, the value of such elements is limited to a range between $-2^{i}+1$ and $2^{i}-1$. In the dot-product module, these input values are first multiplied and stored in {\tt temp[][]}, which requires $2i$ bits to avoid an overflow and maintain precision. Then, the sum of $d$ {\tt temp[][:]} values are stored in {\tt dot\_product[]} which requires $log_2(d) + 2i$ bits. Then, in the exponent computation module, the value of the maximum elements in {\tt dot\_product[]} is subtracted from all elements in {\tt dot\_product[]}, which requires one extra integer bits for {\tt dot\_product[]}. After the exponent computation, the value of {\tt score[]} is now limited to a range between 0 and 1 and thus no integer bit is required. For {\tt expsum}, which is a sum of $n$ values in {\tt score[]}, $log_2(n)$ bits are required, and for {\tt weight[]}, no integer bit is required since their value is still bounded to a range between 0 and 1. For the final {\tt output[]}, $i+log_2(n)$ bits are required.

\para{Fraction Bitwidths} Fraction bitwidth is directly related to the precision of the value. Assume we utilize $f$ (e.g., $f=4$ is used for our evaluation) bits to represent the elements of both inputs (i.e., key matrix and query matrix). In the dot-product module, inputs with $f$ fractional bits are multiplied and stored in {\tt temp[][]}, which now requires $2f$ bits to maintain precision. Since additions (or subtractions) do not change the required number of fractional bits, {\tt dot\_product[]} value also keeps $2f$ fraction bits. Keeping $2f$ fraction bits across this exponent computation also does not lose precision\footnote{When the quantization error is positive (i.e., $\epsilon > 0$), $~e^{x+\epsilon} - e^x = e^{x+\epsilon} (1-e^{-\epsilon}) < \epsilon$ holds true because $e^{x+\epsilon} < 1$ ($\because x < 0$ indicates $x+\epsilon < 0$) and $1-e^{-\epsilon} < \epsilon$.  When $\epsilon < 0$, a quantization error $ e^x - e^{x+\epsilon} = e^x\cdot(1-e^\epsilon) < -\epsilon=|\epsilon|$ because $e^x <1 (\because x\leq 0)$ and $1-e^\epsilon < -\epsilon$. These inequalities prove that error becomes smaller after the exponential function when the exponent part is negative.} and thus {\tt score[]} uses $2f$ fraction bits as well. {\tt weight[]} also uses $2f$ fraction bits since division does not require additional precision as long as the divisor (i.e., {\tt expsum} in our case) is larger than 1. Lastly, {\tt output} requires $3f = 2f$ (for weights) $+ f$ (for values) fraction bits.
 
\begin{figure*}[t]
  \centering
  \includegraphics[width=0.85\textwidth] {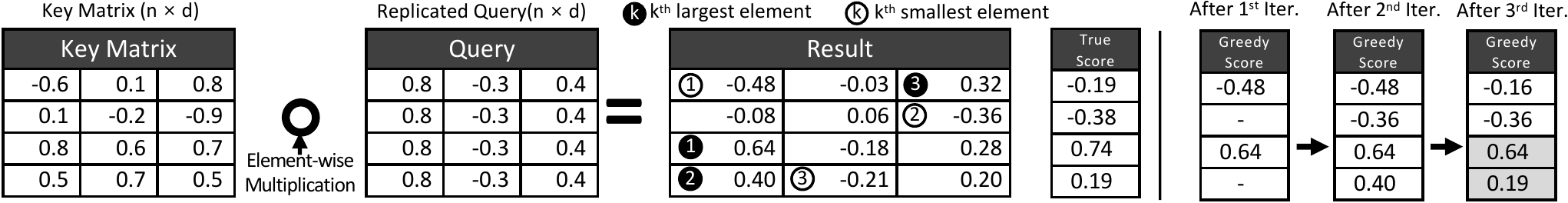}
  \caption{Illustration of the base greedy candidate search algorithm.}
  \label{fig:greedyalg}
\end{figure*}
\subsection{Design Details} 

\para{Offloading Mechanism} \name can be integrated with most devices including CPUs, GPUs, or hardware accelerators for deep learning. Since the base \name is a relatively straight-forward accelerator, it can be integrated to any level of the memory hierarchy. Before invoking \namens, a key matrix and a value matrix should first be copied to the SRAM buffer of \namens. Note that the time it takes to copy these matrices is often not a part of the query response time. For example, in question-answering s, a key matrix and value matrix are ready on {\it comprehension time} (i.e., reads and memorize knowledge) rather than a {\it query response time} (i.e., reads a question and try to find a relevant knowledge to answer the given question). In other words, a key matrix and a value matrix are copied beforehand and the only communication time included to the query response time is the time it takes for a host device to copy a query vector to \namens. Once the query arrives, \name will buffer it to the query queue. Whenever a dot-product module becomes available, it can start the computation for the query. At the end of the computation, the output vector will be buffered to the output queue for the host. 

\para{Use of Multiple \name Units} Most NN processing tasks have a large amount of parallelism. In such workloads, it is often desirable to handle multiple, independent attention computations in parallel. In such a case, it is possible to use multiple copies of our \name units for a different key, value matrices sets. On the other hand, there are often cases where multiple queries are processed to the same set of key and value matrices. In that case, queries can execute in parallel through pipelining in a single \name unit. Note that it is also possible to utilize multiple instances of \name units for the same set of key, value matrices to increase the throughput. 

\para{Choice of $n$ and $d$} \name can be synthesized for different $n$ and $d$ values depending on the needs. In our evaluation, we set $n$ to 320 and $d$ to 64 to fit the largest task we evaluated. However, \name can still handle a task that requires even larger $n$ values. When a larger $n$ is desired, we store first $n$ vectors to the SRAM while leaving other vectors to the DRAM. Since \name accesses both the key matrix and the value matrix in a sequential manner, it is possible to utilize a prefetcher to read them from a memory without exposing memory latency. Note that the size of the key matrix and the value matrix for one of the largest existing models utilizing attention mechanism (i.e., $n=320$ and $d=64$ on BERT model) is still small enough to fit in SRAM so it is likely that the use of DRAM won't be necessary for the near future. Unlike $n$, $d$ is not likely to vary widely since a choice of too large $d$ can lead to decrease in model accuracy \cite{wordembed1, wordembed2}.  For this reason, it often makes sense to simply assume a large enough $d$ and use zero-padding when smaller $d$ is desired. 
%When a larger $d$ is required, we simply utilize multiple memory rows to store information about the one vector. For example, if our set of SRAMs is configured to support up to 64-dimensional vector when we need to store 128-dimensional vector, the first 64 elements of the vector will be stored in a row of the SRAM and the remaining 64 elements of the vector will be stored in the next row of the SRAM. On the other hand, 

\section{Approximate Attention} \label{sec:approximation}
\subsection{Overview} \label{sec:spoverview}

As explained in Section \ref{sec:motiv}, the attention mechanism is essentially a content-based approximate search. In a conventional attention mechanism, the algorithm computes the similarity score between the query vector and all candidates (i.e., each row in the key matrix) and translates scores to weights using softmax function. With those weights, a weighted sum of each row in the value matrix is computed and returned. 

Here, one important point is that most of those weight values are often near-zero. Softmax function is a soft (i.e., differentiable) version of the argmax function which amplifies the value differences between a few large entries and other smaller entries. Since score values are transformed to weights with this function, candidates with relatively small score values get near-zero weight. In addition, these near-zero weights often do not contribute to the accuracy of the model. Rather, it is more of a byproduct of utilizing a differentiable version of the max function, which is important for training, but not for inference. So for these near-zero weights, it is actually more beneficial to treat them as zeros and avoid including them for softmax computation and the following weighted sum computation. 

An even better way is to avoid computing the score at all for rows of the key value matrix that will end up with the low score, and have near-zero weights after the softmax computation. For this purpose, we present an approximation algorithm which can select the candidates that are likely to have a high score without actually computing the score.  The main intuition behind this approximation approach is that it is possible to preprocess the key matrix without affecting the critical path. As explained in Section \ref{sec:motiv}, the key matrix (and the value matrix) is obtained at {\it knowledge comprehension time} rather than {\it question answering time}. By preprocessing the key matrix, our algorithm tries to reduce the number of operations that can be handled at {\it question answering time} where the query becomes available. In addition, on models like BERT where multiple queries (e.g., 320) utilize this same key matrix, the cost of the preprocessing key matrix is amortized and incurs only a limited amount of overhead. 

\subsection{Base Greedy Candidate Search} \label{sec:greedy}

Inner product computation between a query vector and a row in a key matrix is a sum of component-wise multiplication result across $d$-dimensions. The main idea behind our proposed scheme is that looking at a single component-wise multiplication result provides information about the final outcome. More specifically, our proposed algorithm assumes the following: if a multiplication result of a particular dimension is a large positive number, it is likely that the sum of multiplication results for all dimensions (i.e., dot product result) is large as well. Similarly, if a multiplication result of a particular dimension is a large negative number, it is likely that the dot product result is not a large positive value. In fact, a similar intuition is used for other application domains such as information retrieval \cite{greedymips}. 

Figure~\ref{fig:greedyalg} illustrates the basic idea of our algorithm. Given a key matrix and a query vector, this algorithm first replicates query vector across rows to make a replicated query matrix. Then, an element-wise multiplication of these two matrices is performed. Then, an element at the $i$th row and the $j$th column in the resulting matrix represents a $j$th dimension multiplication result between the $i$th row and the query vector. Naturally, the sum of all elements in a single row computes the inner product between a row in the key matrix and the query vector (represented as {\it True Score} in Figure~\ref{fig:greedyalg}).  

The main idea of this algorithm is to look at the largest (or the smallest) element in this result matrix in an iterative way. During a $k$th iteration, the algorithm checks the $k$th largest (and the $k$th smallest) element and adds such value to the corresponding row in the greedy score array. This process is repeated for $M$ (i.e., a user-configurable parameter) times. Once these iterations finish, a greedy score array approximates the true score array. If a row has a positive greedy score, this indicates that the row has one or more relatively large positive components. On the other hand, a row with a negative greedy score indicates that this row has one or more relatively large negative components. Based on this observation, our algorithm selects rows that have positive scores at the end of the iteration as candidates and passes them to the base \name dot product module.

This algorithm, in its current form, has a time complexity of $O(nd~\log~nd)$. In order to select the $k$th largest element from the result matrix, all elements of the matrix should be sorted to avoid performing a linear search every time. Naturally, an $O(nd \log nd)$ time algorithm is not very useful when full dot product computation (i.e., true score computation) takes $O(nd)$ time. Below, we introduce the efficient implementation of this algorithm which exploits preprocessing steps to make the {\it query response time} (i.e., the time it takes from query arrival to the output) not dependent on $n$. 

\subsection{Efficient Greedy Candidate Search} \label{sec:greedyplus}

\begin{figure}[!ht]
  \centering
  \includegraphics[width=0.85\linewidth]{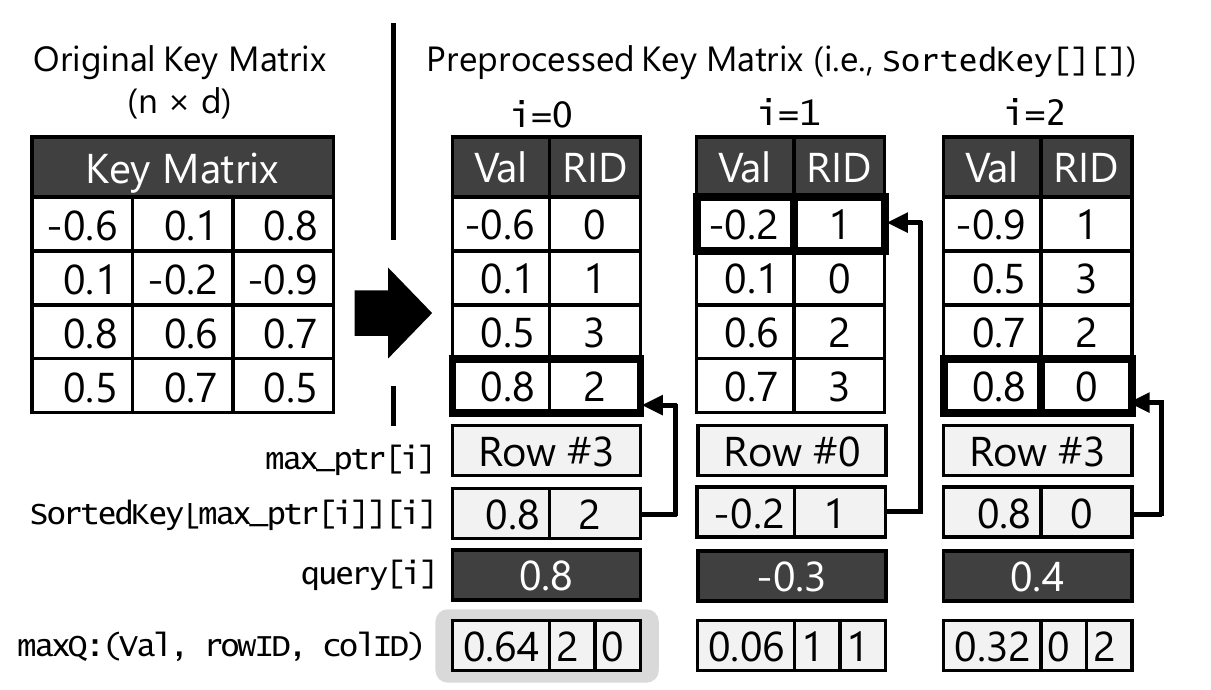} 
  \caption{Illustration of the data structures for the efficient greedy search algorithm. {\tt minQ} operations are omitted for conciseness.}
  \label{fig:datastruct}
\end{figure}

Figure \ref{fig:spalgorithm} shows an algorithm that is functionally identical to the one explained in the previous subsection and Figure \ref{fig:datastruct} shows example data structures utilized in this algorithm. While this algorithm is functionally identical to the one presented in the previous subsection, it utilizes a preprocessing to minimize the latency at the {\it query response time}. In this algorithm, preprocessing step (Line 1-5) only requires the key matrix, which is often available at {\it comprehension time}, where an algorithm comprehends knowledge, past states, etc. for future uses. On the other hand, {\tt candidate\_selection} happens on {\it query response} time where a query is ready. This is often a critical path since it is desirable to answer the provided query in a short time.  Note that there are some models where preprocessing cannot happen off-critical path (e.g., Google BERT). However, models like Google BERT and Transformer utilize self-attention mechanism which utilizes the same key matrix for multiple queries (e.g., 320 times). In such a case, the cost of the preprocessing is amortized so that models' performance can still benefit from approximation. 

\begin{figure}[!th]
\begin{minipage}[b]{1.02\hsize}
\begin{lstlisting}[escapechar=\^,style=myStyle,mathescape=true]
void $\textbf{preprocess}$ (float key[][]) {
  for i = 0 to d:
    sortedKey[:][i] = sort(key[:][i])
    //sortedKey:Sorted List of (val, rowID) pairs
}
int[] $\textbf{candidate\_selection}$ (float query[]) {
  maxQ = MaxPriorityQueue() 
  // MaxPriorityQueue: Priority queue of (val, rowID, colID) tuples
  /* Initialize Pointer */
  for i = 0 to d:
    max_ptr[i] = (n-1 if query[i] > 0 else 0)
  /* Initialize Priority Queue */
  for i = 0 to d:
    entry = sortedKey[max_ptr[i]][i]
    score = entry.val * query[i]
    maxQ.push(score, entry.rowID, i)
  /* Iterative Candidate Selection */
  for iter = 0 to M:
    maxScore, rowID, colID = maxQ.pop()
    if maxScore > 0:
      greedy_score[rowID] += maxScore
    max_ptr[colID] += (-1 if query[colID] > 0 else 1)
    nextEntry = sortedKey[max_ptr[colID]][colID]
    compMultRes = nextEntry.val * query[colID]
    maxQ.push(compMultRes, nextEntry.rowID, colID)
  /* Update Candidates */ 
  for each (row, score) in greedy_score:
    if score > 0:
      candidates.append(row)
  return candidates
}
\end{lstlisting}
\caption{Pseudocode representation of the efficient greedy candidate search algorithm. Note that statements related to {\tt minQ} are omitted for conciseness since they are mostly symmetric to {\tt maxQ} operations.}
\label{fig:spalgorithm}
\end{minipage}
\end{figure}

\para{Preprocessing} During a preprocessing step, each column of the key matrix is sorted and the result is stored in {\tt sortedKey} (Line 1-5 in Figure \ref{fig:spalgorithm}). Then, once the query becomes available, {\tt candidate\_selection} starts. At the beginning of  {\tt candidate\_selection}, the {\tt max\_ptr} (and {\tt min\_ptr}) is initialized for each column. The {\tt max\_ptr} is set to the row index of the entry with the largest value in the column of the {\tt sortedKey} matrix if the query component for the corresponding column (i.e., {\tt query[i]}) is positive. Otherwise, it is set to the entry with the smallest value in the column. The {\tt min\_ptr} is also initialized in a similar, but in the opposite way. Then, priority queues (i.e., {\tt maxQ, minQ}) are initialized. Each entry in each column of the {\tt sortedKey} pointed by {\tt max\_ptr} (and {\tt min\_ptr}) is first multiplied with the corresponding query component and then inserted to the {\tt maxQ} (and {\tt minQ}, respectively) along with its {\tt rowID} and {\tt colID} (Line 12-16). 

\para{Iterative Candidate Selection} After the preprocessing, the iterative candidate selection (Line 17-25) begins. First, an entry from the {\tt maxQ} (and {\tt minQ}) is popped. This entry is the largest (or the smallest in case of {\tt minQ}) entry among the ones currently pointed by {\tt max\_ptr} (and {\tt min\_ptr}). Then, the score value of the entry pointed by {\tt max\_ptr} (and {\tt min\_ptr}) is added to the {\tt greedy\_score} array if it is positive (negative). After this, the {\tt max\_ptr} (and {\tt min\_ptr}) is updated so that it can point the next largest (or the smallest) entry in the column. Finally, the new entry pointed by the updated {\tt max\_ptr} (and {\tt min\_ptr}) is inserted to the respective priority queue. This step is repeated for $M$ times and then rows with positive greedy scores are selected as candidates. Lastly, we apply a small heuristic --- skipping the minQ operation when the cumulative sum of entries selected by {\tt max\_ptr} and {\tt min\_ptr} so far is negative --- to avoid selecting too few candidates when overall similarity scores are low. 

Unlike in the previous version of the algorithm, the complexity of {\tt candidate\_selection} is $M\log d$ ($M$ loops each with $log~d$ complexity from the iterative candidate selection step). This re-structured algorithm now can select likely candidates with a time complexity that scales with user-defined parameter $M$. In practice, to maintain reasonable accuracy, $M$ needs to increase as $N$ increases. However, an important benefit here is that this algorithm provides a user a hyperparameter to adjust the balance between the performance and the accuracy. We evaluate how this algorithm can effectively estimate the set of likely candidates across different $M$ in Section \ref{sec:eval:accuracy}.  

\subsection{Post-scoring Approximation} \label{sec:postscoring}

Once a subset of rows of the key matrix is chosen as candidates, their full scores (i.e., the dot product between the key matrix row and the query) are computed. Then, those scores are used as an input for the softmax function and the outcome of softmax functions are used as weights for the final weighted sum computation. As briefly explained in Section \ref{sec:spoverview}, most of the relatively small dot-product values become near-zero after the softmax and thus have minimal impact on the final outcome. To minimize the computation required to calculate the softmax and the weighted sum, it is beneficial to avoid including some of the low-scoring candidates for those steps. One way to perform an approximation on this step is to sort candidates based on their dot product results and then only including top scoring rows for the next steps. 

One important design choice there is to choose the number of top scoring rows to include for the next step. An easy way is to include a static, predefined number of top scoring rows but a static choice of such number does not necessarily work for all cases. For example, if high-scoring rows form a distribution with very low variance, it is better to include all of those rows for the following softmax computation and the weighted sum computation. On the other hand, if there is only one high-scoring row with many low-scoring rows, it is not beneficial to include more low-scoring rows for the softmax and the weighted sum computation. For this reason, we utilize a dynamic post-scoring approximation scheme which decides whether to include a row for the next steps. Basically, a score for a particular row is compared with the top-scoring row's score and if their difference is larger than threshold $t$, such row is excluded for the next steps. If a row's score is smaller than the top row's score by more than $t$, this means that this row will have a post-softmax weight that is at least $e^t \times$ smaller than that of the top-scoring row. This is because softmax functions utilize the current score as an exponent term of the base $e \approx 2.718$. Throughout the paper, we use $T = 100 * (1/e^t)$ instead of directly using $t$. With this notation, $T$ indicates that an entry should have a post-softmax weight that is at least $T$(\%) of the maximum weight to be included for the next steps.

\section{\namet with Approximation} \label{sec:approxhw}

%It is possible to utilize the proposed approximation scheme introduced in Section \ref{sec:approximation} in software. In fact, the use of approximate similarity search algorithms is popular in the information retrieval community for a significant speedup it can provide. However, we find that the use of approximation algorithms (implemented in software) often results in a substantial slowdown compared to the baseline when $n$ is not very large as in attention mechanisms (e.g., 320 in Google BERT). In such cases, the proposed approximation algorithm cannot fully utilize the conventional computing device due to its irregular data access patterns and data structures.  To address this challenge, we design new specialized hardware accelerator modules for candidate selection and post-scoring approximation to enable the proposed approximation scheme to be efficiently processed. Then, we connect them to the base attention mechanism accelerator introduced in Section \ref{sec:design} to complete \name with approximation capability.

To efficiently implement the approximation scheme introduced in Section \ref{sec:approximation}, we design new hardware accelerator modules for candidate selection and post-scoring approximation. Then, we connect them to the base attention mechanism accelerator introduced in Section \ref{sec:design} to complete \name with approximation capability. 

\subsection{Candidate Selection Module} \label{sec:candidateselection}

\begin{figure}[t]
  \centering
  \includegraphics[width=0.85\linewidth] {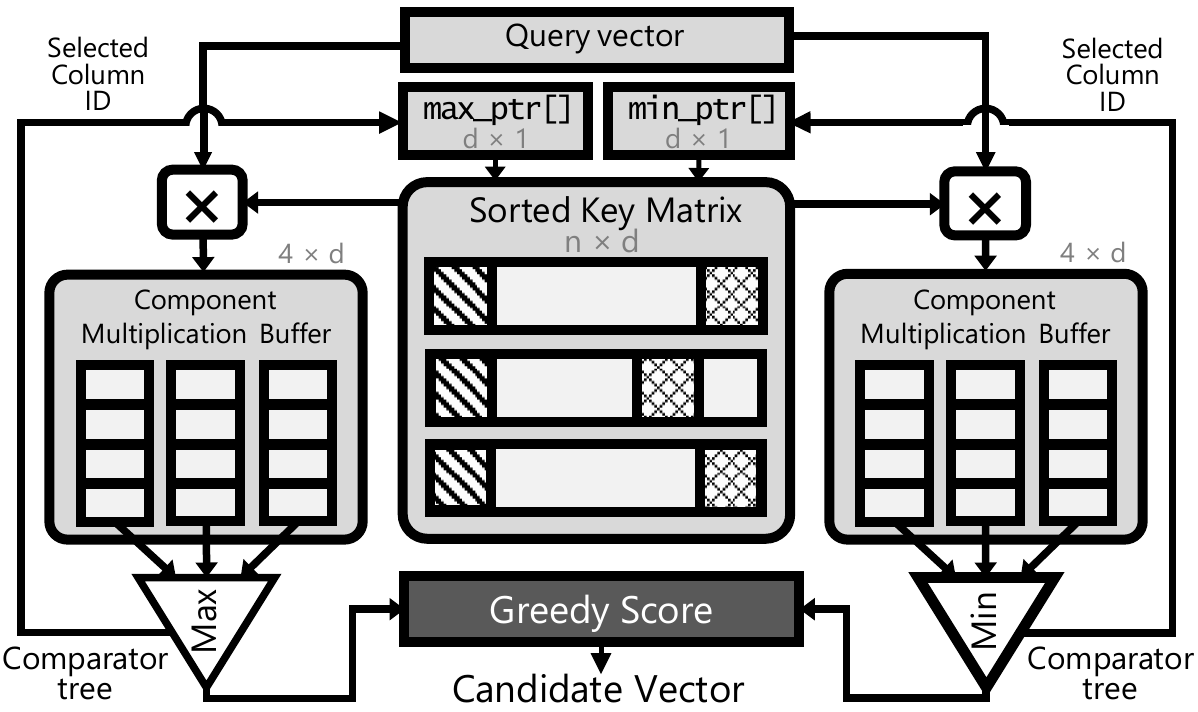}
  \vspace{-0.04in}
  \caption{Simplified block diagram of the \name candidate selection module}
  \label{fig:approxblock}
\end{figure}

This hardware is designed to accelerate the algorithm described in Figure \ref{fig:spalgorithm}. By utilizing the customized hardware and exploiting hardware parallelism, our candidate selection module enables us to significantly reduce the execution time and improve the energy efficiency of the algorithm. Figure \ref{fig:approxblock} shows the block diagram for the candidate selection module. 

This module has SRAM structures which buffer a preprocessed version of the key matrix as described in Section \ref{sec:approximation}. In these SRAM modules,  the sorted version of each column in the original key matrix are stored. And along with the values, each word in the SRAM also includes the row index of the corresponding value in the original key matrix just like in Figure~\ref{fig:datastruct}. It also has two sets of $d$ registers ({\tt max\_ptr} and {\tt min\_ptr}), two multipliers, two sets of $d$ circular queues to buffer component multiplication results for each column, two $d$-dimensional comparator tree to find the maximum and minimum value from $d$ component multiplication results, and a set of greedy score registers where greedy scores are updated.    

\para{Pipeline Design} A naive conversion of the algorithm in Figure~\ref{fig:spalgorithm} to hardware would result in low throughput as the iterative candidate selection (Line 17-25 in Figure \ref{fig:spalgorithm}) has dependency across iterations. Specifically, Line 19 is dependent on Line 25 of the previous iteration and this effectively makes each iteration of the loop to execute in series. We break this backward dependency by pre-executing component multiplication (Line 24) for all $d$ dimensions for the first few iterations in advance. Assuming the critical path of the loop body takes $c$ cycles (e.g., $c=4$ in our implementation), our hardware is initialized with $c \times d$ pre-computed component multiplication results ($c$ items each for each column) in the component multiplication buffer, which consists of $d$ circular queues (Figure~\ref{fig:approxblock}). In steady state, our hardware simply removes one item from this component multiplication buffer (from the column that has the largest component multiplication result) and adds one back to that column $c$ cycles later. Note that this $c$-cycle refill path is fully pipelined to allow our hardware to initiate a new iteration every cycle, thus achieving the throughput of one iteration per cycle. Below, we explain the operation of the pipeline in detail.

\para{Initialization} Our hardware first initializes each of {\tt max\_ptr} and {\tt min\_ptr} to the appropriate values (i.e., either $n-1$ or 0) and reads them from SRAM simultaneously. Then, to fill the component multiplication buffer, our hardware performs i) $2d$ multiplications where each multiply operation multiplies a component of the sorted key matrix pointed by {\tt max\_ptr} (and {\tt min\_ptr}) and the corresponding element of the query, ii) update of {\tt max\_ptr} (and {\tt min\_ptr}) register for each column. This process is repeated for a total of 4 times to fill the two sets of $4 \times d$ buffer. This process requires $8d$ multiplications, and it will normally take $4d$ cycles since this module only has two multipliers. However, to reduce this cost, we use $d$ multipliers in the dot product module and $d$ multipliers in the output computation module of the base \name (Figure~\ref{fig:hwblock}) just for this process. Once this finishes, this module goes to a steady state where only two multiplications happen every cycle. 

\para{Candidate Selection} In steady state, the hardware performs an iterative candidate selection (similar to Line 19-26 in Figure~\ref{fig:spalgorithm}). Every cycle, this hardware performs multiple operations simultaneously: i) load an item from the specific column of the sorted key matrix that is pointed by the corresponding {\tt max\_ptr} (and {\tt min\_ptr}) register and buffer it near a multiplier, ii) perform a multiplication of the buffered data and the corresponding query elements and update the corresponding circular queue of the component multiplication result buffer, iii) find the maximum value among $d$ items (i.e., the oldest items of each circular queue) with a $d$-way comparator and signal the selected column to update {\tt max\_ptr} (and {\tt min\_ptr}), and iv) update the greedy score register array with the value outputs from iii). Note that our candidate selection hardware utilizes a $d$-dimensional comparator tree to obtain the maximum (and minimum) entry among $d$ elements in a single cycle instead of $log d$ cycles. With this hardware support, the complexity of the algorithm in Figure~\ref{fig:spalgorithm} becomes $O(M)$ instead of $O(M log d)$. Since this module performs one iteration every cycle in steady state, it takes $M$ (plus a few extra) cycles to complete the iterative candidate selection part. Once completed, the hardware linearly scans the {\tt greedy\_score} registers (up to 16 contiguous entries per cycle) and sends one row ID (with positive score) to the next module.

\subsection{Post-Scoring Selection Module} 

The post-scoring selection module is in charge of identifying a certain number of top entries from the dot-product results following the mechanism described in Section~\ref{sec:postscoring}. This hardware is integrated at the beginning of the exponent computation module in base \name and its primary function is to select a candidate row with the maximum dot product value among remaining ones. For this purpose, this module simply computes the difference between the maximum dot product value and remaining entries at high throughput (e.g., 16 entries per cycle). If the difference between a compared value and the top-scoring entry's value is larger than the preset threshold $t$, this value is passed to the next module (i.e., exponent computation module). This module essentially consists of 16 subtractors and comparators.

\subsection{\namet HW with Approximation Support}

\begin{figure}[t]
  \centering
  \includegraphics[width=\linewidth]{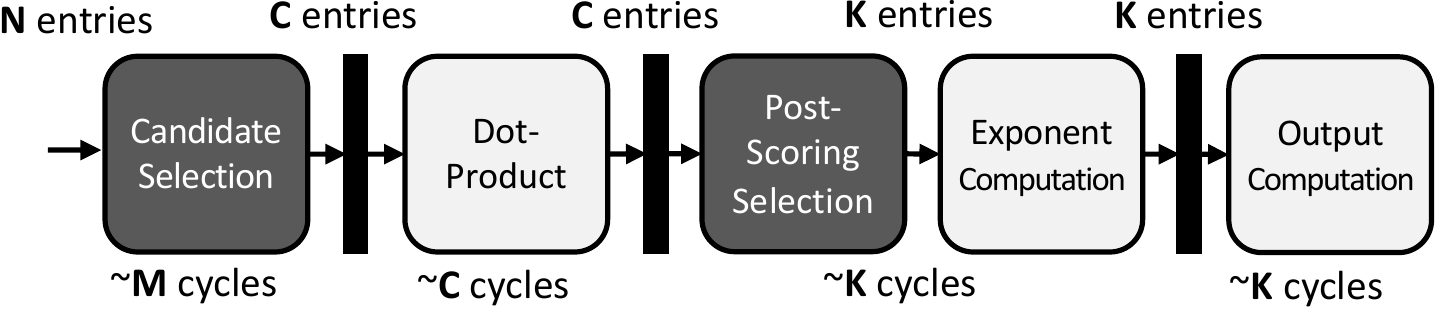} 
  \caption{High-level block diagram of the \name design with approximation.}
  \label{fig:finaldesign}
\end{figure}

Figure \ref{fig:finaldesign} shows the high-level block diagram of the \name design. At the very beginning, given a preprocessed key matrix, the candidate selector module extracts the list of rows and passes it to the dot product module. The dot-product module computes the dot product results for the provided list of candidates and passes it to the post-scoring selector, which is integrated with the exponent computation module. The post-scoring selection module selects a few rows to be included for the exponent computation and then the exponent computation module computes the exponent for those rows. Finally, the output computation module generates the output by computing the weighted sum. Assuming $C$ candidates are selected from the candidate selector module in $M$ iteration and $K$ entries are again selected from the post-scoring selection module, the total latency for \name is $M+C+K+K+\alpha$ cycles where $\alpha$ is a constant. The throughput is limited by the candidate selector module ($\approx M$ cycles) since our candidate selector module selects less than $M$ candidates because i) each iteration may update the greedy score for the same row and ii) only the rows ended up with the positive greedy scores are selected.

\section{Evaluation}
\subsection{Workloads} \label{sec:workload}
We use three representative neural network (NN) models utilizing attention mechanisms. First, we evaluate our \name with Facebook's End-to-End Memory Network (MemN2N)~\cite{memn2n} running bAbI QA task~\cite{babiqa}  which consists of twenty types of question-answering tasks. For each task, a set of statements is provided and the model aims to find the right answer through attention mechanisms which find the most relevant statement to the question. Second, we evaluate our design with Key-Value Memory Network (KV-MemN2N)~\cite{memkvnn} running Wikimovies~\cite{web:wikimovies} question-answering tasks. In this workload, Key-Value Memory Network model first comprehends multiple excerpts about movies from Wikipedia, and then is expected to answer questions about movies. Lastly, we evaluate our proposal with Google BERT (base)~\cite{bert}, which utilizes a self-attention mechanism in Google Transformer \cite{transformer} to solve many tasks in natural language processing. Among the many tasks that this model can handle, we evaluate Stanford Question Answering Dataset task (SQuAD v1.1~\cite{squad}). 

We used the embedding dimension $d=64$ for all workloads but each workload has different $n$. bAbI QA is a relatively small task whose average $n$ (i.e., number of statements for a query) across all test inputs is 20 and the maximum is 50.  For Wikimovies dataset, average $n$ (i.e., the number of potentially relevant knowledge) is 186. Lastly, for SQuAD workload, $n$ (i.e., the maximum length of an input passage and a question in terms of word counts) is 320.

\subsection{Accuracy Evaluation} \label{sec:eval:accuracy}

\para{Methodology} To estimate the impact on the accuracy of our approximation scheme proposed in Section~\ref{sec:approximation}, we implement a software model for approximation and integrate this model with our target workload's official (or endorsed) open-source implementations. Specifically, we use the native Python implementation \cite{web:memnn} for End-to-End Memory Network, Torch implementation \cite{web:memkvnn} for Key-Value Memory network, and Tensorflow implementation \cite{web:bert} for BERT. Note that we only apply approximation techniques for an inference, which is our target, and we use test set inputs for accuracy measurements. For accuracy metric, we utilize one of the main metric used in the relevant paper for the task: accuracy for bAbI QA, Mean Average Precision (MAP) for Wikimovies dataset, and F1 score for SQuAD. 

\begin{figure}[!th]
  \centering
  \includegraphics[width=\linewidth, trim={11pt 11pt 11pt 9pt},clip]{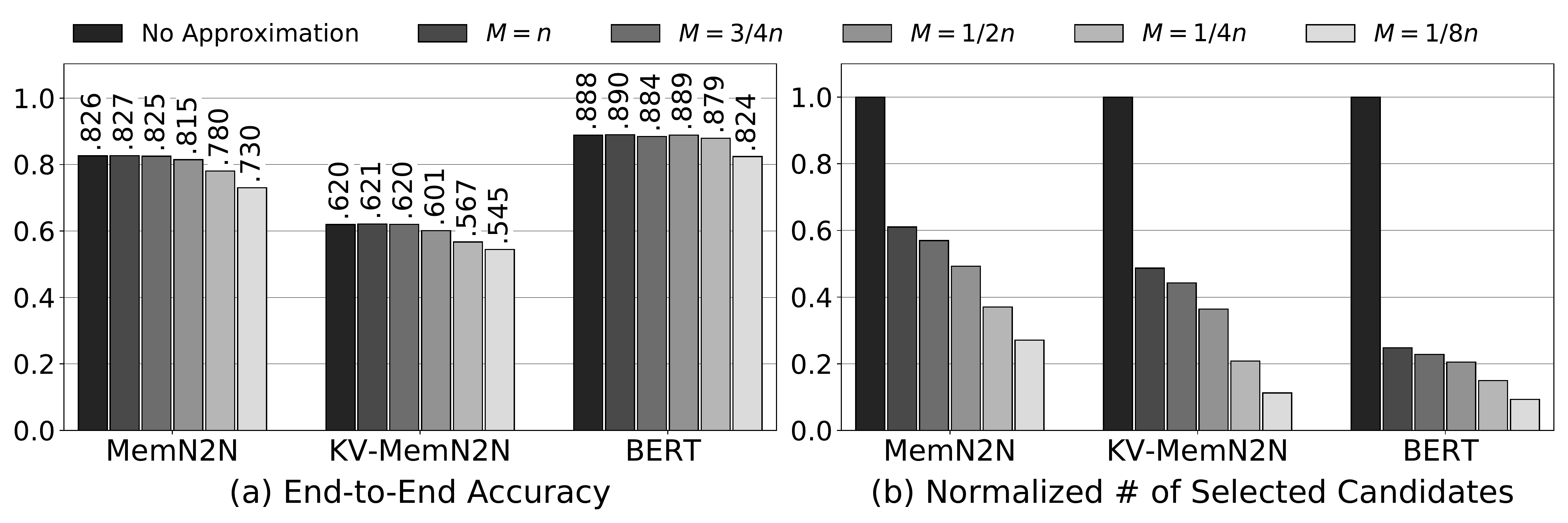} 
  \caption{Impact of proposed candidate selection schemes on accuracy across varying iteration counts.}
  \label{fig:eval1}
\end{figure}

\para{Impact of Candidate Selection} Figure~\ref{fig:eval1}a shows the percentage differences of accuracy metric for three workloads after applying the proposed candidate selection scheme in Section~\ref{sec:greedyplus}. Specifically, we vary $M$ (i.e., iteration count for candidate selection algorithm in Section \ref{sec:greedyplus}) in Figure~\ref{fig:eval1} to check how varying $M$ impacts accuracy and the number of candidates selected. As shown in Figure~\ref{fig:eval1}a, varying $M$ from a large number (e.g., $n$) to a smaller number (e.g., $1/8n$) results in a change in model accuracy. This is because varying $M$ results in a different number of selected candidates as shown in Figure~\ref{fig:eval1}b. Naturally, the larger the number of selected candidates, the accuracy of the model increases; however, it loses benefits of approximation with a large number of candidates. 

\begin{figure}[!th]
  \centering
  \includegraphics[width=\linewidth, trim={11pt 11pt 11pt 9pt},clip]{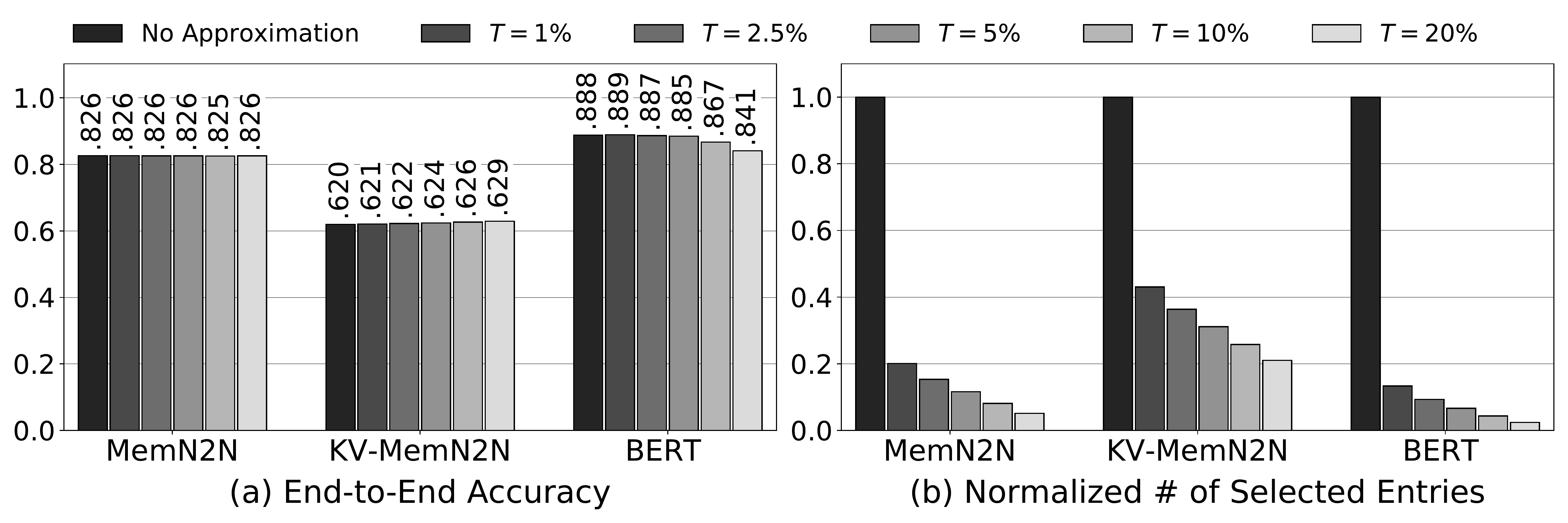}
  \caption{Impact of post-scoring selection schemes on accuracy across varying thresholds.}
  \label{fig:eval2}
\end{figure}

\para{Impact of Post-Scoring Selection} Figure~\ref{fig:eval2}a shows the change in model accuracy with the proposed post-scoring selection scheme (Section~\ref{sec:postscoring}). We vary $T$ (i.e., the threshold for post-scoring selection algorithm) in Figure~\ref{fig:eval2} to identify how $T$ affects to model accuracy. Here, note that an entry is not included for the computation if its post-softmax score would be less than $T$\% of the maximal value. Thus, the lower $T$ indicates more conservative approximation and the higher $T$ indicates more aggressive approximation. As shown in the figure, relatively high $T$ (e.g., 10\%) can still achieve decent accuracy. This essentially proves our assumption that attention mechanism does not really require all rows to be inspected and any row that would end up with low weight can safely be ignored. Figure~\ref{fig:eval2}b shows the normalized number of entries selected in the post-scoring selection scheme. Higher $T$ results in a lower number of selected entries. 

\begin{figure}[!ht]
  \centering
  \includegraphics[width=\linewidth, trim={11pt 11pt 11pt 9pt},clip]{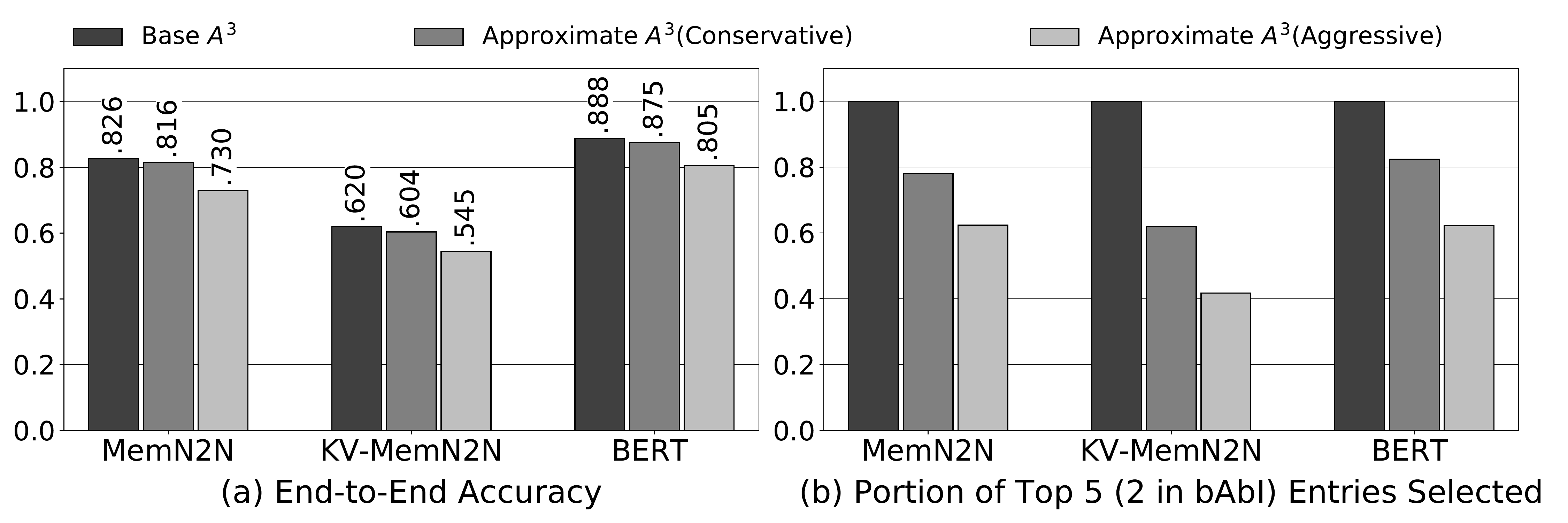} 
  \caption{Impact of the approximation scheme on model accuracy across varying workloads.}
  \label{fig:eval3}
\end{figure}

\para{Impact of Approximation Scheme} Figure~\ref{fig:eval3}a shows the accuracy change after applying both the proposed candidate selection schemes and the post-scoring selection scheme and Figure~\ref{fig:eval3}b shows the portion of true top 2 (bAbI) or top 5 (Wikimovies, SQuAD) entries included after approximation. Here, we evaluate two configurations of our approximation schemes. Approximate (conservative) scheme represents a conservative scheme (with $M=1/2n$ and $T=5$\%), which loses relatively low accuracy (around 1\%) but results in a larger selection size during candidate selection and post-scoring selection. On the other hand, approximate (aggressive) scheme represents an aggressive scheme (with $M=1/8n$ and $T=10$\%) loses an extra accuracy (around 8\%) but results in a much smaller selection size during candidate selection and post-scoring selection. One of the main strengths of our approach is that $M$ and $T$ are configurable. By changing $M$ and $T$, a user always can select the degree of approximation and choose the trade-offs between accuracy and performance/energy efficiency. Figure~\ref{fig:eval3}b shows that more aggressive approximation tends to miss some of the true top 2 (bAbI) or top 5 (Wikimovies, SQuAD) entries compared to the conservative scheme. Note that the aggressive approximation configuration may not be practical on its own for its relatively high decrease in accuracy. In that sense, this configuration is mostly for exploratory purposes. However, one thing to note is that speedups and energy efficiency improvements approximation provide can be translated to improvements in accuracy through the use of larger models. For example, Google BERT has a larger pre-trained model with better accuracy, which naturally spends more time on attention mechanism. The same time/resource-accuracy trade-off is observed in top-performing image classification networks (e.g., Amoebanet\cite{amoebanet}, NasNet\cite{nasnet}) as well. 

\para{Impact of Quantization Scheme} To identify the impact of precision in number representation, we quantize the original input to have $f$ fractional bits (as explained in Section~\ref{sec:quantization}) and then measure its impact on the pipeline. Since our hardware pipeline is carefully designed to avoid precision loss across pipelines, no extra precision loss happens throughout the pipeline. Our experiments show that maintaining a very small number of fractional bits (e.g., $f=4$-bits) has a negligible impact (i.e., less than 0.1\% degradation) on accuracy across all workloads. This proves our assumption that attention mechanism is tolerant to approximation.

\subsection{Performance Results} \label{sec:eval:perf}

\para{Methodology} We implement a cycle-level simulator for our proposed accelerator (running at 1GHz) and integrate it into the open-source implementations of our target workloads to evaluate our accelerator's performance. For comparison, we also profile throughput for attention mechanism processing on Intel Xeon Gold 6128 CPU \cite{intelxeon} (used for all workloads) and NVIDIA Titan V (Volta) \cite{nvidiatitan}  (only used for BERT since other two workloads did not have a GPU implementation). For CPU and GPU measurements, we tried our best to optimize its throughput following Intel performance optimization guidelines for deep learning workloads \cite{inteldeep1, inteldeep2}. 

\begin{figure}[!th]
  \centering
  \includegraphics[width=\linewidth, trim={11pt 11pt 11pt 9pt},clip]{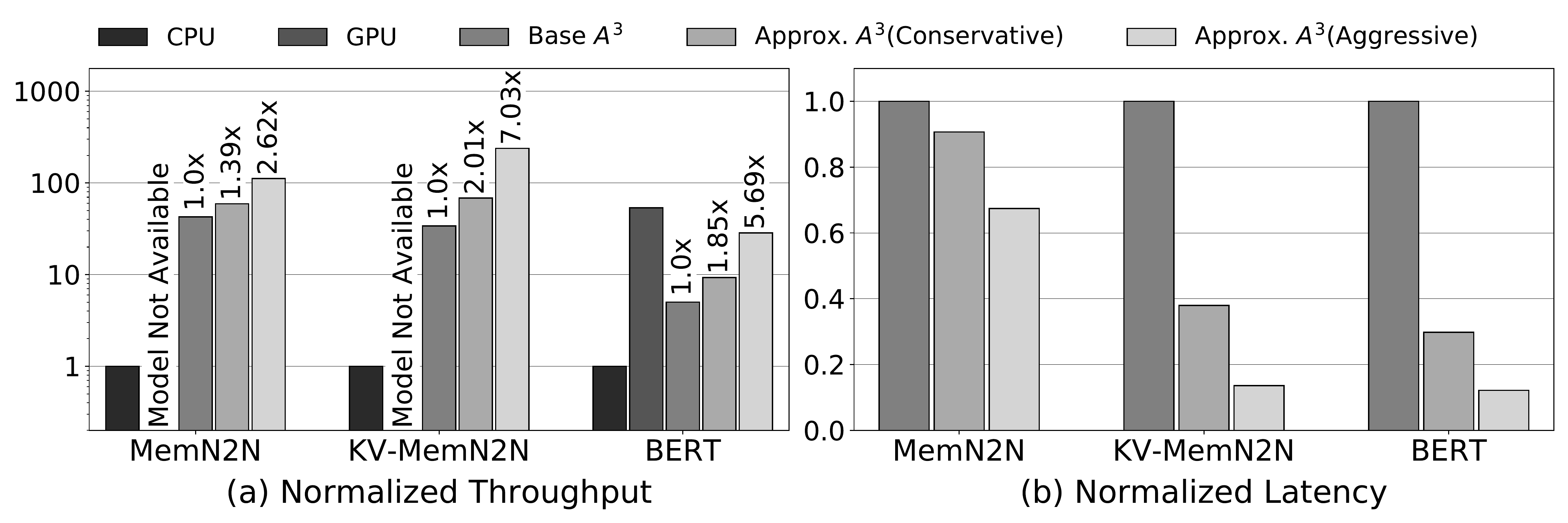} 
  \caption[Normalized average throughput/latency of an attention operation for each workload across platforms.]{Normalized average throughput/latency of an attention operation for each workload across platforms.\footnotemark}
  \label{fig:eval4}
\end{figure}
% n=4 minimum
\para{Throughput} Figure~\ref{fig:eval4}a shows the average throughput of processing an attention operation in the base \name and the approximate \namens, Intel Xeon CPU, and NVIDIA GPU. For each workload, throughput is normalized to that of the CPU. As shown in Figure \ref{fig:eval4}a, both base \name and approximate \name achieve orders of magnitude higher throughput than Intel Xeon CPU on MemN2N and KV-MemN2N. For BERT, \name achives lower throughput than GPU since BERT's self-attention mechanism --- essentially a batch matrix-matrix multiplication instead of a single matrix-vector multiplication --- has easy-to-exploit parallelism. However, since a single \name unit consumes multiple orders of magnitudes less energy than the CPU or the GPU (e.g., $10^3$-$10^4\times$), it is possible to utilize multiple copies of \name for better throughput. Specifically, since BERT's self-attention mechanism has easy-to-exploit parallelism, using multiple \name units can achieve near-perfect scaling behavior. For this reason, it is possible to achieve better throughput than the state-of-the-art GPU by utilizing 6-7 units of approximate (conservative) \name together. This is very surprising result considering that \name is much smaller than GPUs; this is partly because a large GPU often cannot fully utilize its resources for attention mechanism computation whose matrix size is small and the amount of parallelism is less than what GPU can sustain. 

Comparing the base \name and two configurations of the approximate \name show that approximation enables a larger throughput as well. Note that the throughput increase from approximation is low on MemN2N because of its relatively smaller $n$. This can potentially be addressed by utilizing different numbers of modules for each type of module (e.g., use a larger number of candidate selector modules).
\footnotetext{We show the numbers above each bar, which represent the value normalized to the base \name instead of CPU to help readers see the impact of approximation.}

\para{Latency} Figure~\ref{fig:eval4}b shows the average latency of processing an attention operation in the base \name and the approximate \namens. For each workload, latency is normalized to that of the base \namens.  Figure \ref{fig:eval4} shows both base \name and configurations for the approximate \name achieve a very low average latency. Furthermore, both approximation-enabled configurations of \name achieve significantly better latency than the base \namens. This is because the approximate \name performs a substantially lower number of computations thanks to approximation. Comparing two approximation-enabled configurations shows that aggressive approximation can offer noticeably higher speedup than conservative approximation at a relatively low or moderate accuracy cost shown in Figure~\ref{fig:eval3}.

\para{Preprocessing} For BERT, the preprocessing happens on the critical path and thus we included the amortized preprocessing overhead (measured on GPU) to \name bars (only approximate configurations) in Figure~\ref{fig:eval4}. Specifically, BERT utilizes the self-attention mechanism which reuses the same key matrix for $n$ queries ($n$=320) so the effective overhead per query is only $1/n$ of the total preprocessing time. This overhead translates to 7\% (conservative) or 24\% (aggressive) throughput reduction and <11\% latency increase. As shown in Figure~\ref{fig:eval4}, \name achieves significant benefits from approximation even with the (amortized) preprocessing overhead. 

\subsection{Area, Power, Energy and Test Chip} \label{sec:eval:energy}

\begingroup
\renewcommand*{\arraystretch}{1.15}
\begin{table}[!th]
\caption{Area and Power characteristics of \namens.} \label{tbl:areapower}

\centering
\vspace{0.05in}
{\codesize
\begin{tabular}{P{3.2cm} P{.9cm} P{1.6cm} P{1.3cm}}\Xhline{3\arrayrulewidth}
Module & Area (mm$^2$) & Dynamic Power(mW) & Static Power(mW) \\ \Xhline{1.75\arrayrulewidth}
\multicolumn{4}{c}{\bf Modules for Base \name}\\ \hline
Dot Product & 0.098 & 14.338 & 1.265 \\ 
Exponent Computation & 0.016 & 0.224 & 0.053 \\ 
Output Computation & 0.062 & 50.918 & 0.070 \\ \hline
\multicolumn{4}{c}{\bf Modules for Approximation Support}\\ \hline
Candidate Selection & 0.277 & 19.48 & 5.08 \\ 
Post-Scoring Selection & 0.010 & 2.055 & 0.147 \\ 
\hline
\multicolumn{4}{c}{\bf SRAM Modules}\\ \hline 
Key Matrix (20KB) & 0.350 & 2.901 & 0.987 \\
Value Matrix (20KB) & 0.350 & 2.901 & 0.987 \\ 
Sorted Key Matrix (40KB) & 0.919 & 6.100 & 2.913 \\ \Xhline{1.75\arrayrulewidth}
\multicolumn{4}{c}{\bf Total}\\ \hline
\name & 2.082 & 98.92 & 11.502 \\ \Xhline{3\arrayrulewidth}
\end{tabular}
}
\end{table}
\endgroup

\para{Methodology} We implement \name with hardware construction language Chisel \cite{chisel} and compile it to Verilog, and finished all the functional verifications. Then, we synthesize the Verilog code for the 1GHz clock frequency using Synopsys Design Compiler \cite{sdc}
%, TSMC Memory Compiler, 
and TSMC 40nm standard cell library. We used $n=320$ and $d=64$ to fit our largest workload. For number representation (Section~\ref{sec:quantization}), we use $i=4$ and $f=4$. 

%Intel's Xeon Gold 6128 CPU processor
\para{Area} As shown in Table~\ref{tbl:areapower}, \name utilizes less than 2.082mm$^2$ area. Note that our CPU baseline Skylake-SP Intel Xeon Core with 14nm process uses a die area of 325mm$^2$ \cite{skylakesp}, which is 156$\times$ larger than a single \name unit. Similarly, our baseline GPU Titan V has 815mm$^2$ die size \cite{nvidiavolta} at a 12nm technology node, which is 391$\times$ larger than our \namens. And the effective area difference becomes much larger if we consider that our \name is synthesized at 40nm technology. 

\begin{figure}[!t]
  \centering
  \includegraphics[width=\linewidth]{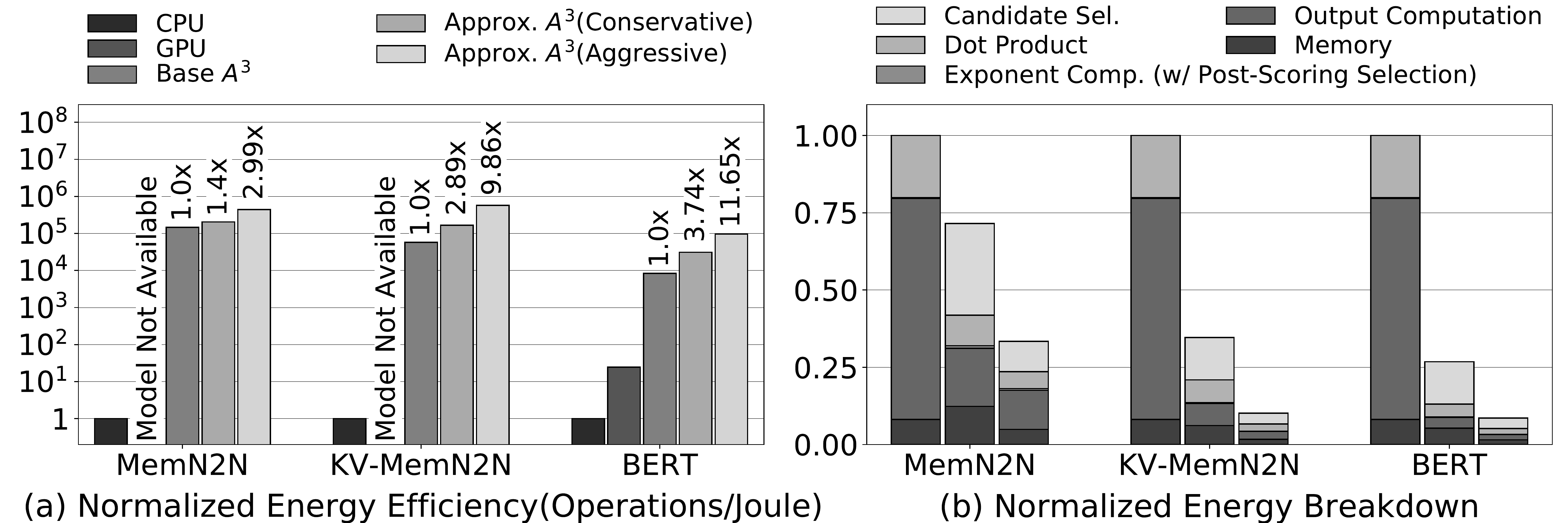} 
  \caption[Energy efficiency and energy breakdown of \name across varying workloads]{Energy efficiency and energy breakdown of \name across workloads \footnotemark[\value{footnote}] For the energy breakdown, three bars represent base \namens, approximate \name (conservative), and approximate \name (aggressive) from left to right.}
  \label{fig:energy}
\end{figure}

\para{Energy and Power} Table~\ref{tbl:areapower} shows the dynamic and static power usage of \name and Figure~\ref{fig:energy} shows the energy efficiency and energy breakdown of our proposed accelerator %(along with CPU and GPU) 
running various workloads. For CPU and GPU energy numbers, we assumed their power consumption is equal to their TDPs. Table~\ref{tbl:areapower} shows that \name spends less than 100mW when all modules are fully utilized. This is already much lower than that of the Intel Xeon CPU (115W TDP) or NVIDIA Titan V GPU (250W TDP). And when running the real workloads, it consumes even less amount of power than its peak power due to a pipeline imbalance resulting from the approximation. As shown in Figure~\ref{fig:energy}a, \namens's low power usage, combined with its high throughput, leads to multiple orders of magnitude energy efficiency improvement compared to the conventional hardware (e.g., over $10^4\times$ on CPU $10^3\times$ on GPU), despite the fact that our accelerators are synthesized at a 40nm technology node. In addition, comparing the base \name and the approximate \name demonstrates that approximation leads to a further energy saving by avoiding unnecessary computation. Lastly, Figure~\ref{fig:energy}b shows that base \name spends most of its energy on the output computation module due to its large register structures. On the other hand, approximate \name spends most energy on candidate selection module because other modules are not heavily utilized once the candidate selection module substantially reduces the number of rows to process. %Note that \namens's power consumption is much (i.e., more than 1000$\times$) lower than that of the Intel Xeon Gold CPU (115W TDP) or the NVIDIA Titan V GPU (250W TDP).

\begin{figure}[!t]
  \centering
  \includegraphics[width=1\linewidth] {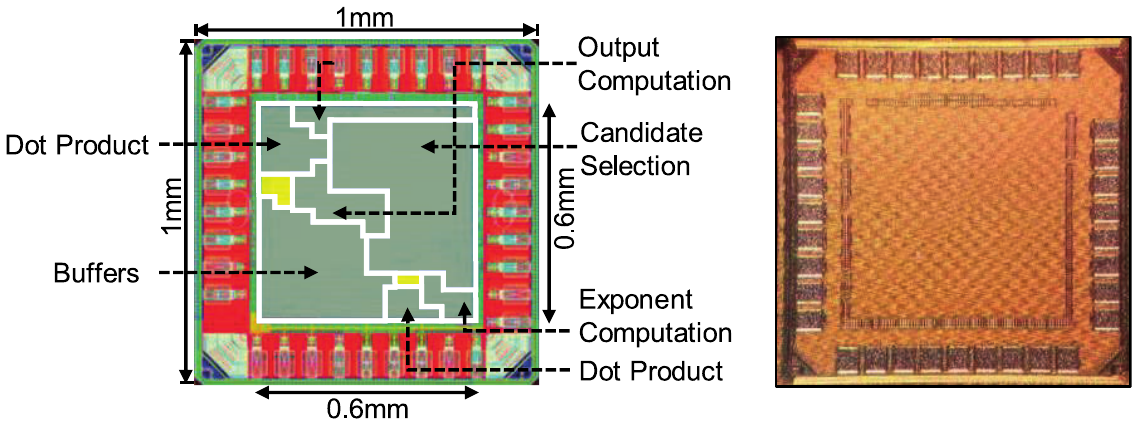}
  \vspace{-0.05in}
  \caption{Post-layout image (left) and die photo (right) of a prototype \name chip} %obatined from Cadence Virtuoso Tool. }
  \label{fig:dphoto}
\end{figure}

\para{Test Chip} We have taped out a scaled-down version of the \name test chip in TSMC 40nm Low Power (LP) technology with standard cells. The test chip implements the full functionality of \name with approximation but is scaled down to fit in a 1mm$^2$ silicon die, including I/O pads, host interface, and other peripherals. The core area is 0.36mm$^2$, which is sized just enough to run the smallest model used for evaluation (MemN2N). Figure~\ref{fig:dphoto} shows a post-layout image and a die photo. The test chip communicates with the host ARMv8 CPU via custom JTAG-like serial interface over 3.3V general-purpose I/O (GPIO) pins. To drive the test chip, we have also written a host device driver as well as a Python-based software testing environment that can run MemN2N. We have verified the functionality of the chip to confirm that everything works as intended.  

% \begin{figure}[!th]
%   \centering
%   \includegraphics[width=\linewidth, trim={11pt 11pt 11pt 9pt},clip]{c4} 
%   \caption{Impact of the quantization on accuracy of the NN models.}
%   \label{fig:quantization}
% \end{figure}

% After the synthesis, we place and route our design and its outcome is shown in Figure~\ref{fig:placeandroute}. 
% \begin{figure}[!th]
%   \centering
%   \includegraphics[draft, width=\linewidth, height=1.2in]{draft} 
%   \caption{Place and Route Result of our design}
%   \label{fig:placeandroute}
% \end{figure}

\section{Related Work}

\para{Attention Mechanism} %Attention mechanism is widely used in many application domains.
%For example, in natural language processing, it is utilized to find the most relevant word or sentence among a large set of potentially relevant entities. 
Attention mechanism in natural language processing is used for translation \cite{transformer, nmt}, question answering \cite{memn2n, bert, memkvnn, attentivereader, bidaf, iarnn}, language inference \cite{decomatt, reasoningaboutentailment}, summarization \cite{rushsummarization, choprasummarization}, document classification \cite{hierarchicalattentionyang, multilinguialpappas}, etc. In addition to natural language processing, the attention mechanism has also been used in computer vision tasks. %For example, the attention mechanism has been used for 
Examples include visual (and multi-modal) question-answering tasks \cite{movieqa, rwmn, pororoqa}, image caption generation \cite{showattend, semanticattention}, image classification \cite{RecurrentModelsofVisualAttention,resnetattention, spatialtransformer, attentioncv}, action recognition \cite{actionrecognitionwithvisualattention}, saliency detection \cite{attentionsaliency}, and so on. Attention mechanism can also be used as a long-term memory mechanism for NNs. Neural Turing Machine \cite{ntm} and Differentiable Neural Computer \cite{dnc} from Google Deepmind focus on this 
%aspect of the attention mechanism 
to enable a NN to solve a complicated task which requires an explicit, long-term memory.  Our work applies to many of these NN models.

\para{Approximate Similarity Search} Similarity search is an important technique in other domains such as recommendation systems. 
%Its computation is very similar to the attention mechanism we described in the paper and thus many of the techniques used for information retrieval are also relevant to our work.
There are several prior works relevant to our work to avoid an exhaustive linear search during a similarity search. For example, some approaches \cite{alsh, cone, clusteringmips} utilize a variant of locality sensitive hashing, a tree-structure, or a clustering algorithm to cluster (or hash) items to different groups before the query arrives.  
%Once the query arrives, it first finds the most relevant cluster (or a hash bucket) then only performs a similarity search within it.
On the other hand, a greedy approach \cite{greedymips} similar to ours performs a greedy iterative search. A few prior works exploit the idea of performing an approximated similarity search for attention mechanisms and implemented them in software \cite{sam, hiearchicalmemory}. 
%Inspired by these software approaches, 
Our work presents a new hardware-friendly algorithm and designs a hardware accelerator to realize the approximation potential in the similarity search. 

%In a recommendation system, a user's preference  and the characteristics of each item is modeled as a vector. Then, the similarity search is performed to find the most relevant item for a given user. 

% \para{Hardware NN Accelerator} 
% %There are an extensive amount of hardware accelerators for NN computations. 
% Various CNN accelerators~\cite{diannao,eyeriss,tpu,dadiannao,shidian,pudian,stripes,brainwave,qiu16,zhang} 
% %for custom ASIC or FPGA device
% and RNN accelerators~\cite{tpu,ese,datelstm,gaornn} are proposed to accelerate NN processing. 
% %by exploiting massive parallelism and optimizing data movements. 
% There is a set of hardware accelerators exploiting sparsity~\cite{ese,eie,cnvlutin,cambricon,scnn} 
% %for a specific context of CNN/RNNs 
% to improve their efficiency; however, these works mostly focus on sparsity (i.e., a large portion of its data is zero) in weights and activation of NNs while our accelerator focuses on the approximation potential of the attention mechanism (which operates on the dense matrix). Some of these accelerators can handle the attention mechanisms as well. 
% %since attention operations can be represented as a dense matrix multiplication, which many of these accelerators are targeted for. 
% However, \name combines the benefit of specialization and approximation to reduce the number of computations significantly, which leads to higher performance and energy efficiency at the same time. 

\para{Hardware NN Accelerator} There exist various hardware accelerators for NN computations. CNN accelerators for custom ASIC or FPGA device~\cite{diannao,eyeriss,tpu,dadiannao,shidian,pudian,stripes,brainwave,qiu16,zhang} and RNN accelerators~\cite{tpu,ese,datelstm,gaornn} are proposed to accelerate NN processing by exploiting massive parallelism and optimizing data movements. There are hardware accelerators exploiting sparsity~\cite{ese,eie,cnvlutin,cambricon,scnn} for a specific context of CNN/RNNs to improve their efficiency; however, these works mostly focus on sparsity (i.e., a large portion of its data is zero) in weights and activation of CNNs or RNNs while our accelerator focuses on the approximation potential of the attention mechanism (operating on a dense matrix). 

%Some hardware NN accelerators can handle the attention mechanisms as well since some portions of attention operations can be represented as a dense matrix multiplication, which many of these accelerators are targeted for. However, such accelerators often lack softmax support and thus have to move data back and forth to the host. Treating attention mechanism as an independent primitive obviates the need for these unnecessary data movements. In addition, a very large accelerator like Google TPU gets severely underutilized for small matrix-vector multiplications utilized in attention mechanism. For example, a single matrix multiply unit (MXU) in TPU has the capability to perform up to 65,536 multiplications simultaneously, but when performing a small (e.g., 320$\times$64 matrix, 64$\times$1 vector) matrix-vector multiplication, it only performs 64 multiplications simultaneously. Some relatively small accelerators (e.g., Eyeriss) can achieve similar matrix-vector multiplication efficiency with base \name design (i.e., almost fully utilize its processing elements). However, approximate \name combines the benefit of specialization and approximation to reduce the number of rows in the matrix significantly, which leads to even higher performance and energy efficiency at the same time. Note that it is also possible to envision integrating the candidate selection and the post-scoring modules to existing DNN accelerators as well. 

\para{Hardware-supported NN Op. Approximation} Several works have applied the approximate computing concept to the neural network operations to reduce the amount of computation. Specifically, SnaPEA~\cite{snapea} exploits the unique characteristics of the neural network convolution operation and presents a hardware accelerator that can benefit from an approximation scheme exploiting such characteristics. Similarly, Raha et al~\cite{rar} presents an RnR (reduce and rank) accelerator that can be used to reduce the energy consumption on certain neural network operations exhibiting the reduce-and-rank pattern through approximation. Finally,  there exist several previous works that focus on accelerating MAC (multiply and accumulate) operations prevalent in neural networks through approximate MAC unit~\cite{rev1, rev2,rev3}.

%V. Akhlaghi, et al. “SnaPEA: Predictive Early Activation for Reducing Computation in Deep Convolutional Neural Networks.” 2018 ACM/IEEE 45th Annual International Symposium on Computer Architecture (ISCA) (2018): %662-673. b. A. Raha, et al. "Energy-Efficient Reduce-and-Rank Using Input-Adaptive Approximations," in IEEE Transactions on Very Large Scale Integration (VLSI) Systems, vol. 25, no. 2, pp. 462-475, Feb. 2017.
%rev3  https://ieeexplore.ieee.org/document/8701880

\section{Conclusion} Neural network (NN) has been a popular target for hardware accelerators for its wide applicability, a large amount of computation, massive parallelism, and static computation pattern. However, the presence of an existing accelerator does not necessarily preclude the need for an another accelerator for NN primitives. In fact, when other NN primitives (e.g., CNNs, RNNs) are optimized, it is critical to accelerate relatively less optimized portion according to Amdahl's Law. Our work identifies the importance of the emerging NN primitive --- attention mechanism --- and accelerates it with software-hardware co-design to achieve orders of magnitude energy efficiency improvement over the conventional hardware.

\section*{Acknowledgments} This work was supported by a research grant from SK Hynix and National Research Foundation of Korea grant funded by the Ministry of Science and ICT (PE Class Heterogeneous High Performance Computer Development, NRF-2016M3C4A7952587, and Nano-Material Technology Development Program, NRF-2016M3A7B4909668), and by IDEC (CAD tools). The authors thank Kwangho Lee, Han-Gon Ko, Soyeong Shin, Yejun Ko, Dongsuk Jeon, and Jintae Kim for their help with troubleshooting our \name test chip design. Jae W. Lee is the corresponding author. 

\bibliographystyle{IEEEtran} 
\bibliography{refs}
\end{document}